\documentclass[12pt]{article}

\usepackage{latexsym}

\begin{document}

\title{Generating  $G_2$--cosmologies with perfect fluid in dilaton 
gravity }

\author{
     Stoytcho S. Yazadjiev \thanks{E-mail: yazad@phys.uni-sofia.bg}\\
{\footnotesize  Department of Theoretical Physics,
                Faculty of Physics, Sofia University,}\\
{\footnotesize  5 James Bourchier Boulevard, Sofia~1164, Bulgaria }\\
}

\date{}

\maketitle

\begin{abstract}
We present a method for generating exact diagonal $G_2$-cosmological 
solutions 
in dilaton gravity coupled to a radiation perfect fluid and with a 
cosmological potential
of a special type. The method is based on the symmetry group of the system 
of $G_2$-field
equations. Several new classes of explicit exact inhomogeneous perfect 
fluid scalar-tensor  cosmologies are presented. 
\end{abstract}


\sloppy

\section{Introduction}

The generalized scalar-tensor theories of gravity are considered as 
the most natural generalization of general relativity. Their importance 
for the current physics is related to the string theory which, in its low 
energy limit,
predicts the existence of a scalar partner of the tensor graviton. A large 
amount 
of research has been devoted to the dilaton cosmology 
\cite{GASPERINI}-\cite{M} (and references therein).

The interest in studying of inhomogeneous (and anisotropic) cosmological 
models is motivated by
the following reasons. As well known, the present universe is not exactly 
spacially
homogeneous, not even at the large scales. Although the homogeneous models 
are 
good approximations of the present universe, there is no reason to assume 
that such
a regular expansion is suitable for description of the early universe. 
Theoretical explanation of the formation of large scale structures in the 
universe also necessitates inhomogeneous models. Contrary to the general 
belief it was shown that the existence of big inhomgeneities in the 
universe does not necessarily lead to an observable effects left over the 
spectrum of CMB \cite{RT}--\cite{AFS}. It was also demonstrated the 
existence of homogeneous but highly anisotropic cosmological models whose 
CMB is exactly isotropic \cite{LNW}. In addition, the inhomogeneous 
cosmological solutions allow us to investigate a number of long standing 
questions regarding the occurrence  of
singularities, the behaviour of spacetime  in vicinity of a singularity 
and the possibility of our universe arising from generic initial data.

In the light of the above reasons the study of inhomogeneous cosmological 
models is necessary and even imperative. The ideal case is to find general 
classes of inhomogeneous cosmological solutions of the field equation 
without any symmetry. However, this seems to be
a hopeless task due to the complexity of the field equations. That is why 
we are forced to assume some simplifications in order to solve the field 
equations. Usually inhomogeneous models with two spacelike commuting 
Killing vectors (the so called $G_{2}$ -cosmologies) are considered. Even 
for these simple cosmological models few exact perfect fluid solutions are 
known in general relativity. The first such class of exact solutions was 
found by Wainwright and Goode \cite{WG}. Other classes were later given in 
\cite{FS}--\cite{D}. All solutions were obtained by assuming the 
separation of variables of the metric components. 

With regard to the scalar-tensor theories, there are no known exact 
inhomogeneous perfect fluid $G_{2}$-cosmological solutions. The reason is 
that the scalar-tensor equations are more complex than the Einstein ones 
and include arbitrary functions of the dilaton field.
That is why finding of exact perfect fluid solutions which hold for all 
scalar-tensor theories is unrealistic in general case. However, for  some 
special equations of state
it is possible to find exact solutions which hold for all scalar-tensor 
theories. In \cite{Yazad1}, methods for generating scalar-tensor stiff 
perfect fluid cosmologies were developed and some explicit solutions were 
presented in \cite{Yazad1}, \cite{Yazad2} and \cite{Yazad3}.

The other equation of state which is realistic and allows us to solve the 
field equations for all scalar-tensor theories (with a special form of the 
dilaton potential) is $\rho=3p$.
It is the purpose of this paper to  present a method for generating 
inhomogeneous perfect fluid diagonal $G_{2}$-cosmologies with equation of 
state  $\rho=3p$ in scalar-tensor theories. As an illustration and 
important consequence of the method, new classes of exact inhomogeneous 
perfect fluid $G_{2}$-cosmological solutions are also  presented for all 
scalar-tensor theories.

\section{Solution generating}

The general form of the extended gravitational action in
scalar-tensor theories is

\begin{eqnarray} \label{JFA}
S = {1\over 16\pi G_{*}} \int d^4x \sqrt{-{\tilde
g}}\left({F(\Phi)\tilde R} - Z(\Phi){\tilde
g}^{\mu\nu}\partial_{\mu}\Phi
\partial_{\nu}\Phi  \right. \nonumber  \\ \left. -2 U(\Phi) \right) +
S_{m}\left[\Psi_{m};{\tilde g}_{\mu\nu}\right] .
\end{eqnarray}

Here, $G_{*}$ is the bare gravitational constant, ${\tilde R}$ is
the Ricci scalar curvature with respect to the space-time metric
${\tilde g}_{\mu\nu}$. The dynamics of the scalar field $\Phi$
depends on the functions $F(\Phi)$, $Z(\Phi)$ and $U(\Phi)$. In
order for the gravitons  to carry positive energy the function
$F(\Phi)$ must be positive. The nonnegativity of the energy of
the dilaton field requires that $2F(\Phi)Z(\Phi) +
3[dF(\Phi)/d\Phi]^2 \ge 0$. The action of matter depends on the
material fields $\Psi_{m}$ and the space-time metric ${\tilde
g}_{\mu\nu}$. It should be noted that the stringy generated
scalar-tensor theories, in general, admit a direct interaction
between the matter fields and the dilaton in the Jordan (string)
frame \cite{GV}. Here we consider the phenomenological
case when the matter action does not involve the dilaton field in
order for the weak equivalence principle to be satisfied. However, the 
method 
we present here holds for the general case since we consider  a radiation 
fluid with
a traceless energy-momentum tensor.   

It is much more convenient from a mathematical point of
view to analyze the scalar-tensor theories with respect to the
conformally  related Einstein frame  given by the metric:

\begin{equation}\label {CONF1}
g_{\mu\nu} = F(\Phi){\tilde g}_{\mu\nu} .
\end{equation}

Further, let us introduce the scalar field $\varphi$ (the so
called dilaton) via the equation

\begin{equation}\label {CONF2}
\left(d\varphi \over d\Phi \right)^2 = {3\over
4}\left({d\ln(F(\Phi))\over d\Phi } \right)^2 + {Z(\Phi)\over 2
F(\Phi)}
\end{equation}

 and define

\begin{equation}\label {CONF3}
{\cal A}(\varphi) = F^{-1/2}(\Phi) \,\,\, ,\nonumber \\
2V(\varphi) = U(\Phi)F^{-2}(\Phi) .
\end{equation}

In the Einstein frame action (\ref{JFA}) takes the form

\begin{eqnarray}
S= {1\over 16\pi G_{*}}\int d^4x \sqrt{-g} \left(R -
2g^{\mu\nu}\partial_{\mu}\varphi \partial_{\nu}\varphi -
4V(\varphi)\right) \nonumber \\ + S_{m}[\Psi_{m}; {\cal
A}^{2}(\varphi)g_{\mu\nu}]
\end{eqnarray}

where $R$ is the Ricci scalar curvature with respect to the
Einstein metric $g_{\mu\nu}$.

The Einstein frame field equations then are

\begin{eqnarray} \label{EFFE}
R_{\mu\nu} - {1\over 2}g_{\mu\nu}R = 8\pi G_{*} T_{\mu\nu}
 + 2\partial_{\mu}\varphi \partial_{\nu}\varphi \nonumber \\  -
g_{\mu\nu}g^{\alpha\beta}\partial_{\alpha}\varphi
\partial_{\beta}\varphi -2V(\varphi)g_{\mu\nu}  \,\,\, ,\nonumber
\end{eqnarray}

\begin{eqnarray}
 \nabla^{\mu}\nabla_{\mu}\varphi = - 4\pi G_{*} \alpha (\varphi)T
+ {dV(\varphi)\over d\varphi} \,\,\, ,
\end{eqnarray}

\begin{eqnarray}
\nabla_{\mu}T^{\mu}_{\nu} = \alpha
(\varphi)T\partial_{\nu}\varphi \,\,\, . \nonumber
 \end{eqnarray}

Here $\alpha(\varphi)= {d\ln({\cal  A}(\varphi))/ d\varphi}$ and
the Einstein frame energy-momentum tensor $T_{\mu\nu}$  is
related to the Jordan frame one ${\tilde T}_{\mu\nu}$ via
$T_{\mu\nu}= {\cal A}^2(\varphi){\tilde T}_{\mu\nu}$. In the case
of a perfect fluid one has

\begin{eqnarray}\label{DPTEJF}
\rho &=&{\cal A}^4(\varphi){\tilde \rho}, \nonumber \\
p&=&{\cal A}^4(\varphi){\tilde p},  \\
u_{\mu}&=& {\cal A}^{-1}(\varphi){\tilde u}_{\mu} . \nonumber
\end{eqnarray}

In the present paper we consider space-times admitting two hypersurface 
and mutually orthogonal Killing vectors $K_{1}={\partial \over 
\partial{y}}$ and  $K_{2}={\partial \over \partial{z}}$. We also require 
the dilaton field to satisfy

\begin{equation}
{\cal L}_{K_{1}}\varphi = {\cal L}_{K_{2}}\varphi = 0 
\end{equation}

where ${\cal L}_{K}$ is the Lie derivative along the Killing vector $K$.

  The metric can be presented in the Einstein-Rosen form

\begin{eqnarray}
ds^2 =D(t,x)[-dt^2 + dx^2] \nonumber \\  + B(t,x)[C(t,x)dy^2 + 
C^{-1}(t,x)dz^2] 
\end{eqnarray}

and the fluid velocity is given by 

\begin{equation}
u = D^{-1/2}{\partial \over \partial t} .
\end{equation}

In what follows we will consider a scalar potential of the form  
$V(\varphi)=\Lambda =const$
(i.e. $U(\Phi)= 2\Lambda F^2(\Phi)$). 

Under all these assumptions we obtain the following system of partial 
differential equations:

\begin{eqnarray}\label{EQ1}
- \partial^2_{t}\ln{D}  + \partial^2_{x}\ln{D} + \partial_{t}\ln{D} 
\partial_{t}\ln{B} -
\partial^2_{t}\ln{B} \nonumber \\ - {\partial^2_{t} B\over B}  + 
\partial_{x}\ln{D} \partial_{x}\ln{B} - (\partial_{t}\ln{C})^2  \nonumber 
\\ = 8\pi G_{*} (\rho + 3p)D  + 
4(\partial_{t}\varphi)^2 -4\Lambda D 
\end{eqnarray}

\begin{eqnarray}
\partial^2_{t}\ln{D}  - \partial^2_{x}\ln{D} + \partial_{t}\ln{D} 
\partial_{t}\ln{B} 
\nonumber \\+ \partial_{x}\ln{D}  \partial_{x}\ln{B}    -  
\partial^2_{x}\ln{B}  -
{\partial^2_{x} {B}\over B } - (\partial_{x}\ln{C})^2  \nonumber \\ = 8\pi 
G_{*} (\rho - p)D + 4(\partial_{x}\varphi)^2  + 4\Lambda D 
\end{eqnarray}

\begin{eqnarray}
 \partial_{t}\ln{B} \partial_{x}\ln{D} + \partial_{t}\ln{D} 
\partial_{x}\ln{B}  + 
 \partial_{t}\ln{B} \partial_{x}\ln{B} \nonumber \\ - 2 
{\partial_{t}\partial_{x}B\over B} -  \partial_{t}\ln{C} 
\partial_{x}\ln{C} = 4\partial_{t} \varphi \partial_{x}\varphi
\end{eqnarray}

\begin{eqnarray}
{\partial^2_{t} B\over B} - {\partial^2_{x} B\over B} = 8\pi G_{*} (\rho - 
p)D + 4\Lambda D
\end{eqnarray}

\begin{eqnarray}
{1\over B} \partial_{t} \left(B \partial_{t}\ln{C}\right) -  {1\over B} 
\partial_{x} \left(B \partial_{x}\ln{C}\right) = 0
\end{eqnarray}

\begin{eqnarray}\label{EQ6}
{1\over B} \partial_{t} \left(B \partial_{t}\varphi \right) -  {1\over B} 
\partial_{x} \left(B \partial_{x} \varphi \right) = 0
\end{eqnarray}

The above system of partial differential
equations (\ref{EQ1})--(\ref{EQ6})  is invariant under the group of 
symmetries $Iso(\mathcal{R}^2)$. Let us introduce 

\begin{equation}
X=\left(%
\begin{array}{c}
  \ln C\\ 2\varphi
\end{array}%
\right) \in  \mathcal{R}^2 .
\end{equation}

The explicit action of the group is given as follows:

\begin{equation}
X \rightarrow MX + \xi
\end{equation}

where $M\in O(2)$ and $\xi \in \mathcal{R}^2$.

The group of symmetries can be used to generate new solutions from known 
ones, especially
to generate solutions with nontrivial dilaton field from pure general 
relativistic  $G_2$-cosmologies.  

The subgroup of translations corresponds to a constant shift of the 
dilaton field
($\varphi \rightarrow \varphi + const$) and to a constant rescaling of the 
metric
function $C$ ($C\rightarrow const\times C $).

That is why, without loss of generality we shall restrict ourselves to the 
subgroup $SO(2) \in Iso( \mathcal{R}^2)$ consisting of the matrixes: 

\begin{equation}
M=\left(%
\begin{array}{cc}
  \cos(\theta) & \sin(\theta) \\
  -\sin(\theta) & \cos(\theta) \\
\end{array}%
\right).
\end{equation}

The remaining discrete $Z_{2}$ subgroup corresponds to the transformations
$C\rightarrow C^{-1}$ or $\varphi \rightarrow - \varphi$. 
 
Let us consider an arbitrary solution of the $G_2$--Einstein 
equations with a radiation perfect fluid  and cosmological term:

\begin{eqnarray}
ds^2_{E} &=& D_{E}(t,x)[-dt^2 + dx^2]  \nonumber \\ &+& 
B_{E}(t,x)[C_{E}(t,x)dy^2 + 
C^{-1}_{E}(t,x)dz^2] \\
\rho_{E}&=&\rho_{E}(t,x) \\
u^\mu_{E}&=& u^\mu_{E}(t,x).
\end{eqnarray}

The $SO(2)$--transformation then generates a new Einstein frame 
scalar-tensor
solution  as follows:

\begin{eqnarray}
ds^2 &=& D_{E}(t,x)[-dt^2 + dx^2]  \nonumber \\ &+& B_{E}(t,x)[C(t,x)dy^2 
+ 
C^{-1}(t,x)dz^2], \\
\rho &=& \rho_{E}, \\
u^\mu &=& u^\mu_{E}\\
\varphi &=& -{1\over 2}\sin(\theta)\ln{C_{E}}, 
\end{eqnarray}

where 

\begin{equation}
\ln C = \cos(\theta) \ln{C_{E}} .
\end{equation}

The $Z_{2}$--transformations can be used to restrict \footnote{When the 
coordinates $y$ and $z$ have the same topology we can restrict the range 
of $\theta$ to $0\le\theta\le\pi/2$ since the metric is invariant under 
the simultaneous  transformations
$C(t,x) \rightarrow C^{-1}(t,x)$ and $y \rightarrow z$. } 
the range of the parameter $\theta$ to 
$0\le\theta\le\pi$. Let us note that for the particular value 
$\theta=\pi/2$ we obtain plane symmetric solutions.   

The Jordan frame solutions are given by

\begin{eqnarray}
F[\Phi(t,x)] &=& {\cal A}^2[-\sin(\theta)\ln C_{E}(t,x)], \\
d{\tilde s}^2 &=& F^{-1}(\Phi)ds^2, \\
{\tilde \rho} &=& F^2(\Phi)\rho_{E}, \\
{\tilde u}^{\mu} &=& F^{-1/2}(\Phi)u^{\mu}_{E} .
\end{eqnarray}

In the above considerations the metric of the $(t,x)$-space was taken to 
be 
in an isotropic form. It should be noted and it is easy to see that 
the solution  generating method is applicable for an arbitrary form
of the $(t,x)$-space metric.

\section{Examples of explicit exact inhomogeneous cosmological solutions}

As an illustration of the solution generating method we consider 
some classes of explicit exact inhomogeneous scalar-tensor cosmologies 
with
$\Lambda=0$.

\subsection{Class 1}

Let us consider Senovilla's solution \cite{SEN} (see also \cite{FS}): 

\begin{eqnarray} 
ds_{E}^2 &=& T^{4}(t) \cosh^{2}(3ax) [-dt^2 + dx^2] \nonumber    \\ &+& 
B_{E}(t,x)[T^3(t)\sinh(3ax)dy^2  \\ &+&  
T^{-3}(t)\sinh^{-1}(3ax)dz^2],\nonumber \\
8\pi G_{*} \rho_{E} &=& 15a^2 T^{-4}(t)\cosh^{-4}(3ax), \\
 u_{E} &=& T^{-2}(t)\cosh^{-1}(3ax){\partial\over \partial t}  
\end{eqnarray}

where 

\begin{eqnarray}
T(t) = \lambda_{1}\cosh(at) + \lambda_{2}\sinh(at), \\
B_{E}(t,x) = T(t)\sinh(3ax)  \cosh^{-2/3}(3ax),
\end{eqnarray}

and $a>0$, $\lambda_{1}$ and $\lambda_{2}$ are arbitrary constants.

The solution generating method gives the following 
scalar-tensor solution:

\begin{eqnarray}
ds^2 &=& T^{4}(t) \cosh^{2}(3ax) [-dt^2 + dx^2]  \nonumber \\ &+& 
B_{E}(t,x)[T^{3\cos(\theta)}(t)\sinh^{\cos(\theta)}(3ax)dy^2  \\  &+& 
T^{-3\cos(\theta)}(t)\sinh^{-\cos(\theta)}(3ax)dz^2 ] ,\nonumber \\
\varphi &=& -{1\over 2}\sin(\theta)\ln[T^{3}(t)\sinh(3ax)]
\end{eqnarray}

\subsection{Class 2}

 Wainwright and Goode's solution \cite{WG} is given by:

 \begin{eqnarray}
  ds^2_{E} &=& \sinh^{4}(2qt)\cosh^{2}(3qx)[-dt^2 + dx^2]  \\
  &+& B_{E}(t,x)[\tanh^{3}(qt)dy^2 + \tanh^{-3}(qt)dz^2 ], \nonumber\\
  8\pi G_{*}\rho_{E} &=&  15q^2 \sinh^{-4}(2qt)\cosh^{-4}(3qt) , \\
   u_{E} &=& \sinh^{-2}(2qt)\cosh^{-1}(3qx) {\partial\over \partial t}.
 \end{eqnarray}

where

\begin{equation}
B_{E}(t,x) = \sinh(2qt)\cosh^{-2/3}(3qx) 
\end{equation}

and $q>0$ is an arbitrary constant.

The corresponding scalar-tensor image of that solution 
is the following:

\begin{eqnarray}
ds^2 &=& \sinh^{4}(2qt)\cosh^{2}(3qx)[-dt^2 + dx^2]  \\
+ &B_{E}(t,x)&[\tanh^{3\cos(\theta)}(qt)dy^2 + 
\tanh^{-3\cos(\theta)}(qt)dz^2 ], \nonumber\\
\varphi(t,x) &=& -{3\over 2}\sin(\theta)\ln[\tanh(qt)] .
\end{eqnarray}

\subsection{Class 3}

The solution found by Davidson is the following \cite{D}:

\begin{eqnarray}
ds^2_{E} &=& -(1 + x^2)^{6/5} dt^2 + t^{4/3} (1+ x^2)^{2/5}dx^2  \nonumber 
\\
&+& B_{E}(t,x) [(tx)dy^2 + (tx)^{-1}dz^2], \\
8\pi G_{*}\rho_{E} &=& {12\over 5}t^{-4/3}(1+ x^2)^{-12/5}, \\
u_{E} &=& (1 + x^2)^{-3/5} {\partial \over \partial t}
  \end{eqnarray}

where

\begin{equation}
B_{E}(t,x) = t^{1/3}(1+x^2)^{-2/5}x .
\end{equation}

Its scalar-tensor image is given by:

\begin{eqnarray}
ds^2 &=& -(1 + x^2)^{6/5} dt^2 + t^{4/3} (1+ x^2)^{2/5}dx^2   \\
&+& B_{E}(t,x)[(tx)^{\cos(\theta)}dy^2 
+(tx)^{-\cos(\theta)}dz^2],\nonumber \\
\varphi(t,x) &=& -{1\over 2}\sin(\theta)\ln(tx).
\end{eqnarray}

\subsection{Class 4}

Here as a seed solution we take Collins's solution of Bianchi type 
$VI_{h}$ \cite{COL}:  

\begin{eqnarray}
ds^2_{E} &=& -d\tau^2 + \tau^2 dx^2  \\ + B_{E}(\tau,x) 
&[&\!\tau^{\sqrt{3}b/2} e^{\sqrt{3}x/2 }dy^2  + \tau^{-\sqrt{3}b/2} 
e^{-\sqrt{3}x/2 }dz^2 ], \nonumber  \\
8\pi G_{*} \rho_{E} &=& {3\over 8} {1-b^2\over \tau^2}, \\
u_{E} &=& {\partial \over \partial \tau}
\end{eqnarray}

where $B_{E}(\tau,x)$ is given by 

\begin{equation}
B_{E}(\tau,x) = \tau^{1/2} e^{bx/2}
\end{equation}

and $0< b< 1$.

The corresponding Einstein frame scalar-tensor solution is the following:

\begin{eqnarray}
ds^2 &=& -d\tau^2 + \tau^2 dx^2    \\ + &B_{E}(\tau,x)& 
[\tau^{\sqrt{3}b\cos(\theta)/2} e^{\sqrt{3}\cos(\theta)x/2 }dy^2   
\nonumber \\ &+& \tau^{-\sqrt{3}b\cos(\theta)/2} 
e^{-\sqrt{3}\cos(\theta)x/2 }dz^2 ] ,\nonumber   \\ 
\varphi(\tau,x) &=& -{\sqrt{3}\over 4}\sin(\theta) \left(b\ln\tau  + 
x\right).
\end{eqnarray}

The Einstein frame metric is homogeneous while the dilaton field is not
constant over the surface of homogeneity. So we have "tilted"  
cosmological solution
in the Einstein frame. The Jordan frame solution, however, is 
inhomogeneous.

We could generate many more examples of exact solutions which  are images 
of the
known $G_{2}$-Einstein cosmologies (see for example the solutions given in 
\cite{RS}, \cite{Kamani} and \cite{KSHMC} ). However, the explicit 
solutions given here are representable and  qualitatively cover the 
general case.

It should be noted that the properties of the found solutions 
in the physical Jordan frame depend strongly on the particular 
scalar-tensor theory
and need a separate investigation.

\section{Conclusion}

In this paper we have presented a simple and effective method for 
generating exact $G_{2}$-cosmologies in scalar-tensor theories with a 
potential of a special form and coupled to perfect fluid with an equation 
of state $\rho=3p$. Several classes of explicit exact solutions have been 
given. These solutions are the only known explicit perfect fluid 
scalar-tensor  $G_{2}$-cosmologies.

It is worth noting that the solutions can be found  by assuming the 
separation of variables of the metric components \cite{Yazad4}. 
However, the way we derived the solutions here is much more elegant and is 
applicable to more general case when  the metric components are not 
separable.

\section*{Acknowledgments}

This work was supported in part by Sofia University Grant No3429.

\end{document}